\documentclass[12pt]{article}

\usepackage{PRIMEarxiv}
\usepackage[utf8]{inputenc} % allow utf-8 input
\usepackage[T1]{fontenc}    % use 8-bit T1 fonts
\usepackage{hyperref}       % hyperlinks
\usepackage{url}            % simple URL typesetting
\usepackage{booktabs}       % professional-quality tables
\usepackage{amsfonts}       % blackboard math symbols
\usepackage{nicefrac}       % compact symbols for 1/2, etc.
\usepackage{microtype}      % microtypography
\usepackage{lipsum}
\usepackage{fancyhdr}       % header
\usepackage{graphicx}       % graphics
\usepackage{subcaption}
\usepackage{float}
\usepackage{amsmath}
\usepackage[T1]{fontenc}
\usepackage[all]{nowidow}
\usepackage{setspace}
\usepackage{multirow}
\usepackage{lineno}
\usepackage[style=ieee, backend=biber]{biblatex}
\newcommand{\mycomment}[1]{}
\graphicspath{{media/}}     % organize your images and other figures under media/ folder

\title{Statistical Distributions for Transient Transport}

\author{
  M. Ross Kunz \thanks{Department of Mathematics $\&$ System Engineering, Florida Institute of Technology, mkunz2011@fit.edu}  \thanks{Catalysis and Transient Kinetics Group, Energy Environment Science $\&$ Technology, Idaho National Laboratory} 
  \And
  Debtanu Maiti \footnotemark[2]
  \And
  Gregory Yablonsky \thanks{Department of Energy, Environmental $\&$ Chemical Engineering, Washington University in St. Louis}
  \And
  Rebecca Fushimi\footnotemark[2]
}

\begin{document}
\doublespacing
\maketitle


\begin{abstract}
    This paper introduces the use of statistical distributions based on transport differential equations for clear distinction of transport modes within transient kinetic experiments. 
    More specifically, novel techniques are developed for the transient data obtained through the Temporal Analysis of Products (TAP) reactor and are applicable to experiments where pulse response into a gas flow is used.
    The methodology allows distinguishing between two domains of diffusion transport in heterogeneous catalytic systems, i.e., Knudsen and non-Knudsen diffusion, using statistical fingerprints, and finding the transition domain. 
    Two distribution parameters were obtained that directly result in coefficients that correspond to the concentration and the rate of transport.
    Using a linear relationship between the rate and concentration coefficients, Knudsen diffusion is revealed when the rate of transport is constant and non-Knudsen diffusion is confirmed when the rate of transport coefficient is a function of the concentration coefficient.
    As a result, accurate transport information is obtained while in the presence of instrument drift or noise while investigating higher pressure pulse responses with complex transport.
    As such, experiments where the influence of gas phase reactions is present can be more directly studied.

\end{abstract}

\section{Introduction}

Statistical analysis for chemical engineering is critical for developing inference based on experimental data. 
This can be challenging not only to address measurement noise and drift, but also determining if assumed governing equations influence an experiment. 
One approach of statistical analysis applied to transient measurements is through the residence time distribution (RTD) developed by Danckwerts \cite{danckwerts1958effect, zwietering1959degree}. 
Simply stated, the RTD is a function that relates the injection of a pulsed tracer to the measured outlet response through statistical distributions for insight of a given process \cite{rodrigues2021residence}. 
Applications of the RTD can be found throughout chemical engineering, e.g., the continuously stirred tank and plug flow reactor, as well as biochemical and nuclear applications \cite{evangelista1969scale, chella1984conversion, choi2004residence, escudie2005hydrodynamic, laquerbe1998identification, klusener2007horizontal, berard2020experimental}. 
However, there exists the challenge of determining the functional form of the RTD such that it sufficiently describes the experimental data while adhering to the assumed physics. 
With respect to transient kinetics, this takes on the form of inference of the types of transport. 
This paper introduces a simple phenomenological model based on the RTD considering two terms for transport, Knudsen diffusion and molecular diffusion. 
This novel form is incorporated into the existing TAP RTD theory to enable a way to study complex gas/surface and gas/gas interactions in non-Knudsen.

The seminal paper, "TAP-2: An Interrogative Kinetics Approach" by Gleaves et al. \cite{gleaves1997tap}, provides the foundation of the RTD associated with the TAP reactor and defines key quantities of interest, such as the rate of transport and reaction. 
To ensure that the transport is separated from the rate of reaction, many techniques have been developed through integral and temporal transformations, but all include the required assumption that the transport consists of only Knudsen diffusion \cite{morgan2017forty, reece2017kinetic, shekhtman1999thin, schuurman2007assessment, redekop2014elucidating, constales2001multi, constales2001multi2, constales2004multi, constales2006multi, constales2017precise, redekop2011procedure, yablonskii1998moment, yablonsky2007procedure}.  
This constraint limits the results to only qualitatively address gas/surface interactions. 
To help relieve the Knudsen constraint, an RTD approach has been previously applied to ideal and non-ideal flow patterns, including transient kinetics via the Temporal Analysis of Products (TAP) reactor, albeit without molecular interactions \cite{yablonsky2009new}. 
Differential equation based models, such as the Dust Gas Model, can describe multiple types of transport but have not been used in practice based on the complexity of calculating an analytical or computational solution \cite{svoboda1993fundamental}. 
More generally, there is no direct method for revealing complex reaction mechanisms under any transient TAP experiment where non-Knudsen transport is present as the time-dependence of the response cannot be directly interpreted due to the confounding effects of transport and reaction. 
Only summary statistics, e.g., conversions or yields calculated from integral quantities, can be determined which drastically reduces the quality of information obtained from the transient experiment.

To make the previously obtained theoretical results more robust to noise and non-Knudsen transport, this paper proposes to introduce statistical techniques, in the form of the RTD, to augment the theoretical form of transport within the TAP reactor. 
This idea is a natural fit as the total number of molecules reaching the end of the reactor at a specific time can be given as a probability based on the transport. 
The concept of probability distributions based on the Fick's Second Law is applied to experimental data to account for non-Knudsen transport and noise. 
As a direct result, the methodology is able to provide a smooth estimate of a transient flux response through extraction of the experimental noise and clear separation of the rate of transport coefficient from concentration.

\section{Methodology}

The objective of this section is to highlight the differences between the potential types of transport that can occur within the TAP reactor, present the mathematical assumptions associated with the TAP theory, and describe how application of distribution techniques can bridge the theory to experimental data. This requires the ability to distinguish the different forms of transport and convert the experimentally observed voltage response to a total number of molecules. 
As a result, this methodology provides a quantitative means of investigating a single TAP flux response to infer the mode of transport and transport velocity without any assumptions of their details. Also, we include a subsection describing the computation of the result such that the need for calibration coefficients and baseline correction is removed.

\subsection{TAP Transport Theory} 

A TAP reactor experiment is ideal for studying diffusional transport/reaction and the residence time distribution due to the experimental setup. A single pulse consists of a mixture of gas species injected into a cylinder moving through a porous medium under vacuum conditions \cite{gleaves1988temporal, gleaves1997tap, shekhtman1999thin, reece2017kinetic, morgan2017forty, kunz2018pulse}. The total number of molecules injected is typically restricted to be within the Knudsen regime, i.e., less than $\sim 10nmol$ depending on the reactor setup \cite{schuurman2007assessment}. It should be emphasized that these constraints result in assuming that gas/gas collisions are considered negligible and gas/solid interactions are strictly first order. This renders the experiment fundamentally different from a more common pulse into gas flow.

Since the molecular inter-dependencies between gas species and non-Knudsen transport are negated within the TAP experiment, the flux $(mol/cm^2/s)$ can be represented as Fick's First Law:
\begin{equation}
\mbox{Flux} = -D \frac{\partial C}{\partial x}
\end{equation}
where $D (cm^2/s)$ is the diffusion coefficient, $C (mol/cm^3)$ is the gas concentration, and $x (cm)$ is the axial coordinate. Note that the diffusion coefficient is assumed to be independent of the concentration when inside the Knudsen regime conditions. As such, Fick's Second Law can be applied to the system to describe the change of concentration based on time and position within the reactor via random motion, i.e., 
\begin{equation}\label{eq:diffusion}
\frac{\partial C}{\partial t} = D \frac{\partial^2 C}{\partial x^2}.
\end{equation}
With initial conditions of
\begin{equation*}
    C = C_0, \quad x < 0, \quad C =0, \quad x > 0, \quad t = 0
\end{equation*}
the solution of Fick's Second Law is given as the Normal (Gaussian) distribution (with dimensionless parameters), i.e.,
\begin{equation}\label{eq:normalficks}
    \frac{C_0}{2\sqrt{2\pi D }} \exp\left(- \frac{x^2 /t}{4D} \right) = \frac{C_0}{ \sqrt{2 \pi \sigma^2}} \exp\left(- \frac{(x/\sqrt{tD} - \mu_\mathcal{N})^2 }{2\sigma_{\mathcal{N}}^2} \right) 
\end{equation}
with an initial concentration of $C_0$, where the solution of Fick's Second Law requires a mean $\mu_\mathcal{N} = 0$ and standard deviation $\sigma_\mathcal{N} = \sqrt{2}$ \cite{crank1979mathematics}. The cumulative distribution function is given as
\begin{equation}\label{eq:cranksum}
 C_0\frac{1}{2} \left[1 - erf \left( \frac{x/\sqrt{tD}}{2}\right) \right]  =  C_0\frac{1}{2} \left[1 - erf \left( \frac{x/\sqrt{tD\epsilon} - \mu_\mathcal{N}}{  \sigma_\mathcal{N} \sqrt{2} }\right) \right]
\end{equation}
where $erf$ is the error function \cite{crank1979mathematics}. The benefit of this form is that the concentration and the flux are given as probability distributions which can be leveraged as a residence time distribution. However, this form assumes that the position is semi-infinite in that the values of $x$ may range from $-\infty$ to $\infty$ which is physically not possible within the TAP reactor as is it assumed no molecules can escape via the pulse valve. Additionally, the Normal distribution is symmetric around the mean which is not the case with respect to the TAP outlet response \cite{gleaves1997tap}. In the semi-infinite case, the Normal distribution can help describe the residence time distribution for the TAP reactor; however, additional boundary conditions must be applied.

\begin{figure}[]
\centering
\includegraphics[width=0.75\linewidth]{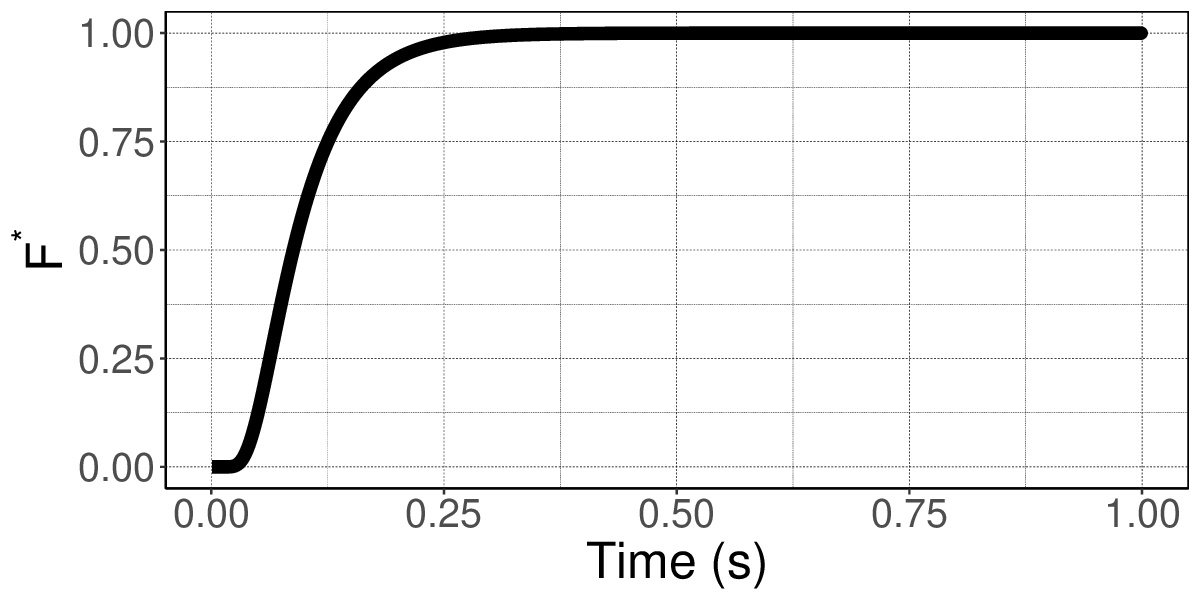}
\caption{Visualizing the properties of the infinite sum within the SDC. Note that monotonically increasing function $F^*$ is clearly asymmetric in time and the domain is naturally restricted to be positive.}
\label{fig:gleavesinfinitesum}
\end{figure}

To account for the reactor geometry and physical requirements, the initial conditions and boundary conditions are updated such that: 
\begin{align*}
    0\leq x \leq L, \quad t > 0,  & \quad C = \delta N / \epsilon A L\\
    x = 0, \quad &  \quad \frac{\partial C}{\partial x} = 0\\
    x = L, \quad & \quad C=0
\end{align*}
where $L (cm)$ is the length of the reactor, $t (s)$ is time, $N (mol)$ is the number of mole, $A (cm^2)$ is the cross-sectional area and $\epsilon$ $($dimensionless$)$ is the bed porosity which is typically assumed to be 0.4. With the given boundary conditions, separation of variables is commonly used to reduce the problem to two ordinary differential equations where the portion with respect to the length has an additional ``forcing'' term based on the non-negativity constraint in $x$. More specifically, the solution is given by Gleaves et al \cite{gleaves1997tap} as a trigonometrical series:
\begin{equation}\label{eq:boundedconc}
    C = \frac{4 C_0}{\pi} \sum_{n= 0}^\infty \frac{1}{2n + 1} \exp\left( - ( 2n + 1)^2 \pi^2 \frac{t D}{x^2} \right) \sin{\frac{(2n + 1) \pi x}{L}}.
\end{equation}
Furthermore, when $x$ is assumed to be at the maximal length $L$, i.e., the outlet, and the concentration is rewritten as flow $F(mol/s)$ via the derivative, the TAP outlet response is given as the standard diffusion curve (SDC) \cite{gleaves1997tap}:
\begin{equation}
    F = SDC(t, \eta) = N \pi \eta \sum_{n = 0}^\infty (-1)^n (2n + 1) \exp \left( -(2n + 1)^2 \frac{\pi^2}{4} \tau \right)
\end{equation}
where $\eta(1/s) = D/\epsilon L^2$ and $\tau = t \eta$ is the dimensionless time. The SDC can be rearranged to be more akin to a probability distribution function through separation of the terms dependent on $n$. That is, 
\begin{align} \label{eq:sdcextend}
\mbox{SDC}(t, \eta) &=  \pi \eta N   \sum_{n=0}^{\infty} (-1)^n (2n +1) \exp \left(- (2n + 1)^2 \pi^2  \tau / 4\right)\\
    &=  \pi \eta N   \sum_{n=0}^{\infty} (-1)^n (2n +1) \exp \left(-  (4n^2 + 4n + 1) \pi^2 \tau /4 \right) \\
    &=   \pi \eta N  \exp \left( -\pi^2 \tau /4 \right)  \sum_{n=0}^{\infty} (-1)^n (2n +1) \exp \left(-   n(n + 1) \pi^2 \tau  \right) \label{eq:sdcpreend} \\
    \label{eq:sdcend} &= \pi \eta N \exp \left( -\frac{\pi^2}{4}\eta t  \right) F^*(t, \eta, n) 
\end{align}
where $F^*$ is a monotonically increasing function with respect to time similar to that of Equation~\ref{eq:cranksum}. Figure~\ref{fig:gleavesinfinitesum} displays $F^*$ with $(t = [0.001, \ldots, 1], \eta = 1, n = 2000)$ such that $F^*$ must be asymmetric and must have a support such that $t\eta$ is strictly greater than zero. Note that when the SDC is broken down in this manner, the scalar $(\eta \pi^2/4)$ is required to be the same within both exponential functions, i.e., outside and inside $F^*$. In other words, the rate of transport must be equal to the scalar term within $F^*$ due to the imposed Knudsen diffusion and boundary conditions.  Due to the imposed boundary conditions, which reflect the physical requirements, the SDC cannot be described by the Normal distribution and a different distribution must be used as the residence time distribution.

Going back to the boundary conditions can give some hints on what distribution approximates $F^*$. When breaking down the partial differential equation to an ordinary differential equation using separation of variables, a forcing function is added to account for the boundary conditions causing the solution for $x$ to go from a homogeneous to an inhomogeneous system. A way to modify the Normal distribution in Equation~\ref{eq:normalficks} is to have a non-negative support via the natural log of the SDC support, i.e., $\ln(\tau)$. This results in the Lognormal probability distribution function such that:
\begin{equation}
\frac{1}{\tau \sigma \sqrt{2\pi}} \exp \left( -\frac{(\ln (\tau) - \mu)^2}{2\sigma^2} \right) \label{eq:lognormalpdf}
\end{equation}
and the cumulative distribution function is given as:
\begin{equation}\label{eq:lognormalcdf}
F^* \approx \Phi(\tau|\mu, \sigma) = \frac{1}{2}  \left[ 1 + erf\left( \frac{\ln (\tau) - \mu}{\sigma \sqrt{2}} \right) \right]
\end{equation}
where value of $\mu$ is the geometric mean of the distribution and \emph{not} the arithmetic mean as seen in the Normal distribution $(\mu_\mathcal{N})$ and $\sigma$ is the log scale. With respect to the SDC, the distribution parameters $\mu = -\pi^2/4 - \eta$ to reflect the scalar in Equation~\ref{eq:sdcpreend} and $\sigma = 1 / 2$ due to the standard deviation of $F^*$ similar to that of Equation~\ref{eq:cranksum}. Based on the properties of the Lognormal distribution, the mode is given as $\exp(\mu - \sigma^2)$, the mean is $\exp(\mu + \sigma^2/2)$ and the median is $\exp(\mu)$ \cite{casella2002statistical}. Smaller values of $\mu$ forces the median of $F^*$ earlier in time while $\sigma$ controls the dispersion or spread of $F^*$ over time.  Rather than adding an additional function to the homogeneous solution in the form of forcing, the natural log transformation enforces the boundary conditions with a specific $\mu$ and $\sigma$. 

Equation~\ref{eq:lognormalcdf} can also be connected to the cumulative concentration given in Equation~\ref{eq:cranksum} to describe the accumulation of the injected gas as a function of time. The difference between the two equations is based on the log of the distribution support, to deal with the boundary conditions, and Equation~\ref{eq:cranksum} includes an initial concentration coefficient $C_0$. Rather than scaling $F^*$ by mole per volume to convert to concentration, the SDC in Equation~\ref{eq:sdcend} only includes the term $N$ for the injected quantity. Hence, $N F^*$ is the cumulative amount of injected gas observed at the end of the reactor with respect to time. Furthermore, the derivative, given as Equation~\ref{eq:lognormalpdf}, is the amount of injected gas as a function of time where the peak is the mode of the Lognormal distribution. This is an important distinction as $F^*$ will always be a function of the injected quantity while the exponential term in Equation~\ref{eq:sdcend} describes the rate of transport.

Combining the estimation of $F^*$ and the exponential function outside the infinite sum (Equation~\ref{eq:sdcend}), the SDC is a product of two distinct distributions: an exponential probability distribution function $\mbox{Exp}(\cdot)$ with rate parameter $\lambda$ and the Lognormal cumulative distribution function given in Equation~\ref{eq:lognormalcdf}. The product of these two distributions create what we will term as the Generalized Diffusion Curve (GDC), i.e., 
\begin{align}
\lambda \exp \left( - \tau \lambda \right) \Phi(\tau|\mu, \sigma) 
&= \mbox{Exp} \left(\tau |\lambda \right) \Phi(\tau |\mu, \sigma) \label{eq:gdc_dist}\\
&= \beta \mbox{ } GDC(\tau|\lambda, \mu, \sigma) \label{eq:GDC}.
\end{align} 
where $\beta$ is the normalization coefficient such that the area of the flux is equal to one. 
As such, the flux response is a function of three dimensionless parameters: $\lambda$ which is the rate of transport parameter, $\mu$ which is the geometric mean of dimensionless time, while $\sigma$ is the log scale of dimensionless time. 
The distributional form allows for specific calculations based on the moments/summary statistics, e.g., mean, mode, and variance, to determine the separate effects of the rate of transport and injection amount.

\subsection{Fingerprinting Transport via Distribution Properties} \label{sec:distributions}

The SDC has convenient properties as the area under the curve is equal to one, the dimensionless mean residence time is $1/2$, the peak residence time is $1/6$, and the product of the peak residence time and the peak height of the normalized flux is equal to a constant $0.31$ \cite{gleaves1997tap}. This subsection provides a means of distinguishing types of transport, i.e., Knudsen diffusion and molecular diffusion, by examining the parameters $\lambda$ and $\mu$ with respect to the mean residence time through similar means as the SDC. 

Assuming Knudsen diffusion, the parameters of the GDC can be tied directly to the mean residence time of the SDC. 
In the case where $\eta = 1$, the mean residence time $(\tau_{res})$ of the SDC is equal to $1/2$ and the sum of the expected values within the GDC is given as
\begin{equation}\label{eq:gdcmeans}
    \tau_{res} \approx \frac{1}{\lambda} + \exp\left( \mu + \frac{\sigma^2}{2} \right) = \frac{4}{\pi^2} + \exp\left( -\frac{\pi^2}{4} + \frac{(1/2)^2}{2} \right)  = 0.501.
\end{equation}
Or when examining the relationship between $\lambda$ and $\mu$ with respect to the solution of the SDC assuming Knudsen diffusion,
\begin{equation}
\frac{\lambda}{\exp(-\mu)} = \frac{\pi^2}{4}  \exp \left(-\frac{\pi^2}{4}  \right) \approx 0.209. \label{eq:mu2lambda}
\end{equation}
As such, a ratio value above  $0.209$ is the boundary of only Knudsen diffusion based on the ideal Knudsen conditions at $10$ $nmol$ \cite{schuurman2007assessment}. 
This can be generalized by assuming that a series of $\eta$ values can be estimated by a linear combination of $\lambda$ and $\mu$ such that,
\begin{equation}\label{eq:lm_eta}
    \boldsymbol{\eta} = \frac{1}{\boldsymbol{\lambda}} + \exp\left( \boldsymbol{\mu} + \frac{\sigma^2}{2}\right)
\end{equation}
where bolded parameters indicate vectors rather than scalar coefficients. 
Using a series of parameters based on consecutive pulses allows the detection of change in the means with respect to the amount of concentration and hence the type of transport. 
More explicitly, since Knudsen diffusion does not depend on concentration, and hence injection size, $\lambda$ will be \emph{constant} over a series of pulses within the Knudsen regime. 
Beyond the Knudsen regime, we expect a linear relationship between $\lambda$ and $\exp(-\mu)$ as both will change as a function of concentration.
The dependencies between $\lambda$ and $\exp(-\mu)$ allow for the fingerprinting of the different forms of transport as well as quantitatively determining the effects of the rate of transport.

\subsection{Experimental Noise and Computational Issues}

To fit an experimental flux response, the flux will be first be transformed to be dimensionless units. First, the flux is subtracted by its minimum then divided by the total area to ensure that the flux is non-negative and has an area that corresponds to an approximate probability distribution function. Note that this shift in baseline also includes a non-zero mean based on the noise within the flux which is assumed to be approximately Gaussian \cite{roelant2007noise}. Therefore, the fit of an experimental flux response must include an intercept term ($\bar{x}$) to account for the baseline as well as a vector ($\epsilon_n$) to account for the approximately Gaussian noise within the flux, i.e., 
\begin{equation}
\bar{\mbox{Flux}}(t) = \bar{x} + \beta GDC(t \eta|\lambda, \mu, \sigma) + \epsilon_n \label{eq:gdcopt}
\end{equation}
where $\bar{\mbox{Flux}}$ is dimensionless flux. 
The same form of this equation can also be applied to the SDC with a change of parameters to $\eta$.
Note that the equation above can be altered to be a function of time rather than dimensionless time, i.e., $t$ rather than $t\eta$. 
If so, then both $\lambda$ and $\mu$ would be scaled by $\eta$ resulting in units of $1/s$. 
Computationally, each coefficient is fit using the non-linear least squares Levenberg-Marquardt algorithm that can be found within most commonly used analytical software \cite{more2006levenberg}. The values of $\beta$ are the normalizing coefficients for the GDC distribution such that the total area of the flux is equal to one.

Below is an outline of the application of GDC for a single flux response:
\begin{enumerate}
    \item Remove negative values by subtracting by the flux minimum
    \item Area normalize the flux%:
    \item Use non-linear least squares for the optimization of coefficients in Equation~\ref{eq:gdcopt}
    \item Calculate the conversion as $1 - \sum_i \beta_i$ for each distribution.
\end{enumerate}

\section{Results}

In the \textbf{simulated} pure diffusion case, fitting the GDC to a dimensionless SDC with a value of $\eta$ set to $1.0 (1/s)$ resulted with the estimated parameters of $\lambda = 2.46$, $\mu = -2.43$ and $\sigma=0.50$. As such, the corresponding correction from $\mu$ to $\lambda$ given in Equation~\ref{eq:mu2lambda}, was found to be $0.217$ which is slightly above the theoretical $0.209$. However, even with the slight deviations from the theory, the fit resulted in an overall Root Mean Square Error (RMSE) of $0.002$ and an $R^2$ of $0.999$.

\begin{figure}[]
    \centering
    \includegraphics[width=0.75\linewidth]{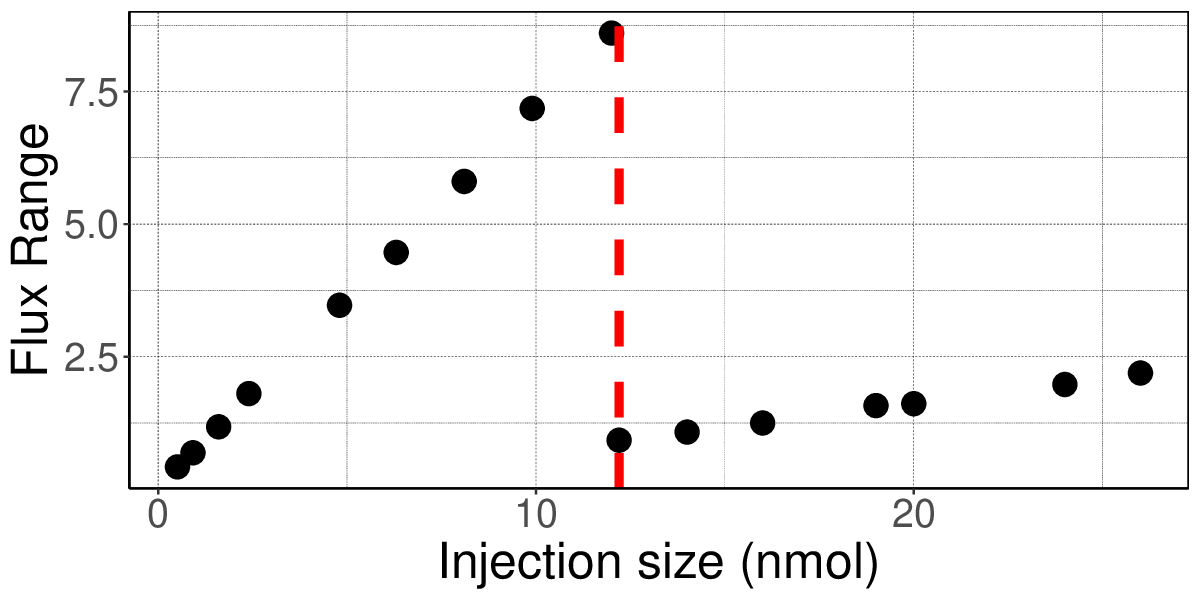}
    \caption{Plot of the flux range per injection size. For each injection size, the flux range increases due to the number of mol injected. There is a transition point at 12.2$nmol$ where the signal amplifier gain setting was descreased to fully measure the flux.  }
    \label{fig:nonknudsenrange}
\end{figure}

To verify the ability of the GDC to fit \textbf{experimental} non-Knudsen transport, the GDC was tested by fitting an argon response over a range of $nmol$ injected into the TAP reactor. 
More specifically, a sequence of experiments was performed at $30^\circ C$ over a range of 0.51 to 24 $nmol$ per pulse where the ideal range for Knudsen diffusion is less than $10$ $nmol$ \cite{shekhtman1999thin}. 
To ensure the whole flux was captured based on the mass spectrometer sensitivity, the gain intensity \footnote[2]{The amplifier gain is used in the conversion of the nanoamp signal from the mass spectrometer to the $(0-10V)$ measuring voltage of the data acquisition system.} was modifed from 8 to 7 at 12.2 $nmol$. 
Figure~\ref{fig:nonknudsenrange} displays the raw data flux range, i.e., the maximum minus the minimum value of each flux, prior to any baseline correction or calibration where the red dashed line indicates the point in which the gain was modified. 
As the pulse injection size increases, there is a increasing trend in the magnitude of the flux for each distinct gain setting. 
This trend is not monotonically increasing within each gain as there are unique baseline and calibration coefficients for each flux response. 
However, the change at $12.2$ $nmol$ causes the flux range to decrease in magnitude for the proper measurement of the flux response.
Such data is difficult to analyze using traditional methods as the trend in the injection size is not consistent due to the scaling of the flux response based on multiple gain settings.
Then, in accordance with our methodology, probability density based methods with either the SDC or GDC must be used as in Equation~\ref{eq:gdcopt}.

\begin{figure}[]
    \centering
    \includegraphics[width=0.75\linewidth]{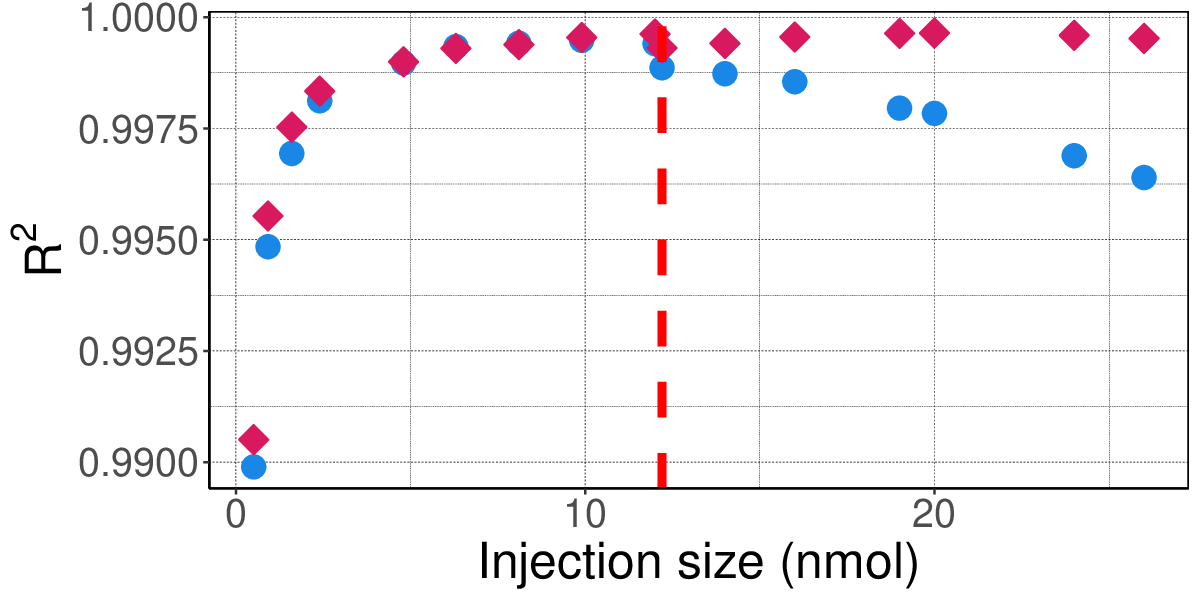}
    \caption{Comparison of the $R^2$ values obtained by the SDC (blue) and GDC (red diamonds) for each injection size. The red dashed line inidcates where the gain setting was changed at $12.2$ $nmol$. The GDC was able to outperform the SDC, as determined by the $R^2$ in all cases but especially when beyond the assumed Knudsen regime of $<10$ $nmol$. }
    \label{fig:nonknudsensamples}
\end{figure}

Both SDC and GDC were fit to each flux response, while accounting for the baseline and scale, where the $R^2$ is displayed in Figure~\ref{fig:nonknudsensamples}.
Both the SDC and GDC performed worse in smaller injection sizes when the noise dominates the signal. 
The difference in the $R^2$ between each method is minimized at an injection size of 9.9 $nmol$, but the GDC outperfomed the SDC in estimation of the flux especially after 10 $nmol$. 
It is interesting to note that the GDC was able to better capture the flux response within assumed Knudsen regime of less than 10 $nmol$.
Over every flux, the average Mean Square Error (MSE) ratio of the SDC over the GDC was approximately 2.7. 
When the total number of molecules injected was more than $\approx 10 nmol$, the SDC underestimates the curve. 
In all cases, the GDC was able to accurately fit each pulse response based on visual inspection.

\begin{figure}[]
    \centering
    \includegraphics[width=.75\linewidth]{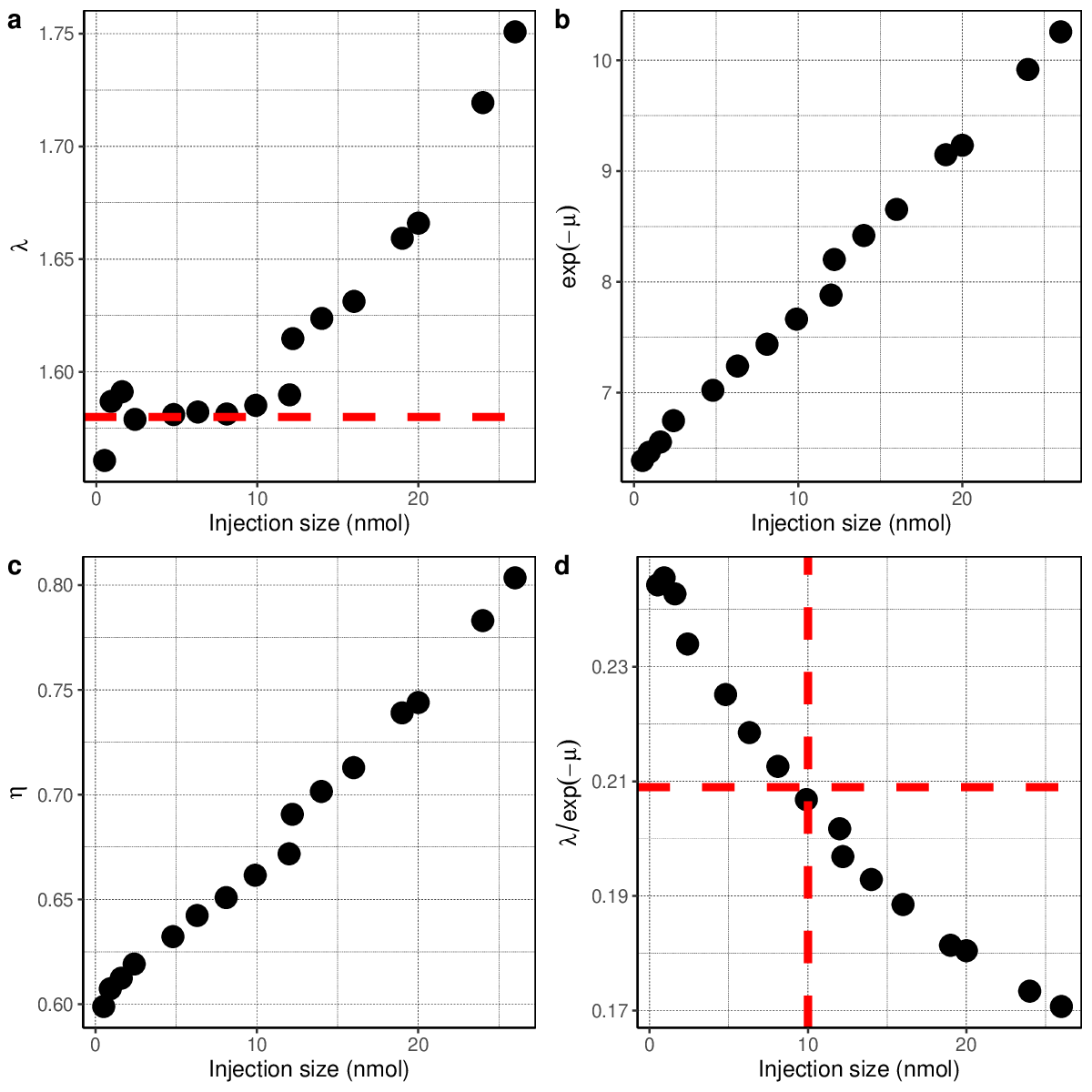}
    \caption{Comparison of the GDC parameters $\lambda$ and $\mu$ fit over varying amounts of injected $nmol$ (a and b). Knudsen diffusion occurs at approximately $\leq 10 nmol$. Sub-figure (c) plots the obtained values of $\eta$ for the SDC which is proportional to the GDC parameter $exp(-\mu)$. Sub-figure (d) compares the ratio $\lambda / exp(-\mu)$ to the ideal indicator of Knudsen diffusion of 0.209 at $10$ $nmol$ which confirms the theoretical derivation.}
    \label{fig:nonknmol}
\end{figure}

To further investigate the GDC fit to each pulse response, the parameters $(\lambda, \mu)$ must be examined over the pulse series. Figure~\ref{fig:nonknmol} compares the $\lambda$ and $\mu$ parameters within the GDC and the value of $\eta$ obtained by the SDC over the series of injected $nmol$. 
Within all cases, the value of $\eta$, estimated by the SDC, shows that the mean residence time will shift linearly, even in the Knudsen regime, over the injected gas quantity. 
This is also the case for $\mu$ as it is always a function of concentration and will shift the peak residence time. 
This trend does not depend on the transport regime whether in Knudsen or non-Knudsen.
The value of $\mu$ determines the point in which $F^*$ reaches a maximum of one (as the total amount of injected substance) and hence approximately where the inflection point occurs as the exponential term forces the flux to approach zero. 
To be concrete, the mode of the Lognormal distribution, i.e., $\exp(\mu - \sigma^2)$ is the inflection point of the gas injection distribution as function of time.
Furthermore, Figure~\ref{fig:nonknmol} (d) validates the theoretical constant of $0.209$ found in Equation~\ref{eq:mu2lambda} at approximately $10$ $nmol$.

\begin{figure}[]
    \centering
    \includegraphics[width=0.75\linewidth]{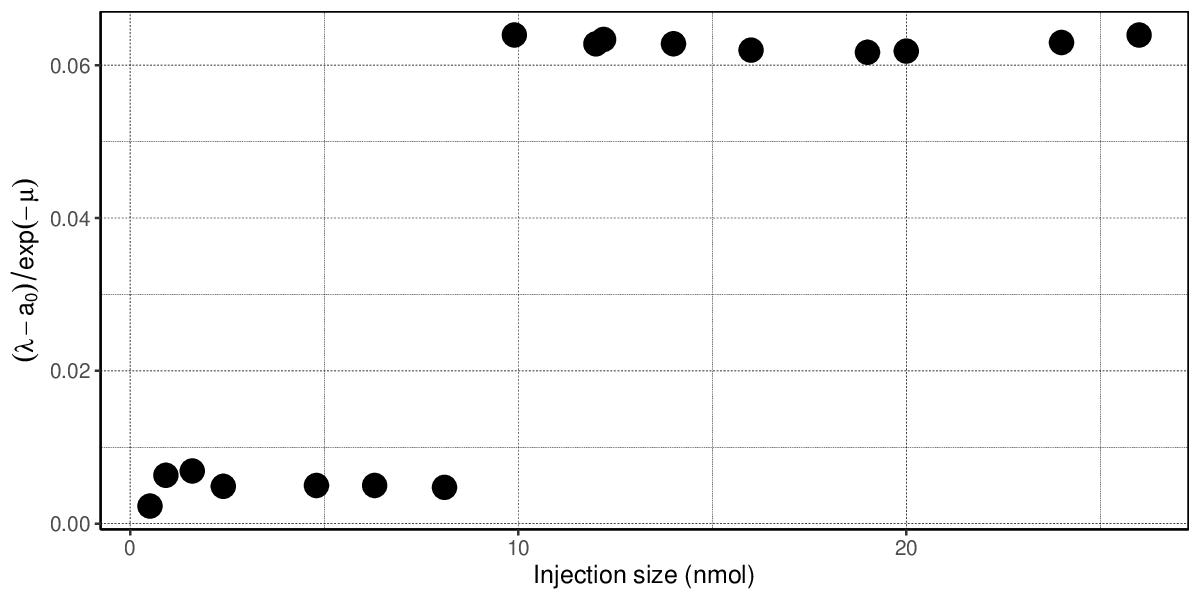}
    \caption{Ratio of $\lambda$, minus the intercept, and $\exp(-\mu)$. The intercept is removed to examine only the effects of concentration. The coefficients below $10$ $nmol$ were found not to be statistically significant based on Table~\ref{tab:coefs}.  } %Comparison of the $R^2$ values obtained by the SDC (blue) / GDC (red diamonds) for each injection size. The red dashed line inidcates where the gain setting was changed at $12.2$ $nmol$. The GDC was able to outperform the SDC, as determined by the $R^2$ in all cases but especially when beyond the assumed Knudsen regime of $<10$ $nmol$. 
    \label{fig:nonknudsenlambda}
\end{figure}

Unlike $\mu$ and $\eta$, the value of $\lambda$ exhibits different trends depending on the type of transport.
At $10$ $nmol$ and less, $\lambda$ exhibits a constant trend signifying the experiment is within the Knudsen diffusion regime as the rate of diffusion $(\lambda)$ is approximately constant. 
This was verified through application of OLS, as mentioned in Section~\ref{sec:distributions}, with the dependent variable as $\lambda$ and independent variable as $\exp(-\mu)$, i.e.,
\begin{equation}
    \boldsymbol{\lambda} = a_0 + a_1 \exp(-\boldsymbol{\mu}) + \boldsymbol{\epsilon}_n \label{eq:lambda_expmu} 
\end{equation}
where $a_0$ is the intercept and $a_1$ is the regression coefficient associated with $\exp(-\mu)$, and $\epsilon_n$ is a vector to account for the noise. 
This equation is the basis for distinguishing different types of transport \footnote[2]{
The value of $\mu$ was defined statistically as the geometric mean of the distribution.
Physico-chemically, $\exp(-\mu)$ can be interpreted as the intensive characteristic of the amount of
substance within the pulse, i.e. the concentration $(C)$ or, in the general case, activity $(a)$. Then, $\mu$
can be understood as the characteristic which is a logarithmic function of concentration/activity,
$\mu = a_0 + a_1 \ln(C) \mbox{ or } \mu = a_0 + a_1 \ln(a)$. In accordance with chemical thermodynamics, the value of $\mu$ presented in this paper can be interpreted as the characteristic
which is similar to the well-known chemical potential. It is interesting to note that in chemical
thermodynamics $\mu$ is the traditional symbol for the chemical potential. Symbolic coincidence
between the presented statistical $\mu$ and thermodynamic $\mu$ is quite surprising.}.
When the coefficient $a_1$ is not statistically significant, then it can be assumed that only Knudsen diffusion occurs as $\lambda$ can be modeled as a constant rather than a function of $\exp(-\mu)$. 
Table~\ref{tab:coefs} (Rows 1 and 2) displays the OLS results when the injection size is less than $10$ $nmol$. 
Assuming a significance threshold of $0.05$, then the $Pr(>|t|)$ for $a_1$ was found not to be statistically significant. 
As such, when less than $10$ $nmol$, $\lambda$ can be estimated solely as the intercept and hence is within the Knudsen regime. 
With an increase in injection amount beyond $10$ $nmol$, the rate parameter $\lambda$ increases with the amount of injection, see Figure~\ref{fig:nonknmol} (a). 
The linear relationship between $\lambda$ and $\exp(-\mu)$ was verified through OLS given in Table~\ref{tab:coefs} (Rows 3 and 4) where both the intercept and slope, i.e., $a_0$ and $a_1$, were found to be statistically significant. 
Furthermore, two OLS models were created with a dependent variable as injection size and independent variables of $\lambda$ and $\exp(-\mu)$ separately, resulted in no statistically significant difference between the two slopes.
Figure~\ref{fig:nonknudsenlambda} displays the ratio of $\lambda$ and $\exp(\mu)$ after accounting for the intercept obtained in the OLS estimate to show that the slope when the injection size is greater than $10$ $nmol$ is approximately constant. 
As such, the pulse series of $\lambda$ beyond $10$ $nmol$ includes Knudsen diffusion and transport as a function of concentration.
In general, we will assume that the results obtained when the pulse injection size is greater than $10$ $nmol$ is a direct consequence of Knudsen diffusion and non-Knudsen transport based on the regression coefficients and their associated $p$-value.

\begin{table}[]
    \centering
    \begin{tabular}{l | l |r|r|r|r}
        & Coef & Estimate & Std. Error & t value & Pr(>|t|)\\
        \hline
        \multirow{2}{*}{$<10$ $nmol$} & $a_0$ & 1.546 & 0.071 & 21.796 & 0.000\\
        &$a_1$ & 0.005 & 0.010 & 0.480 & 0.652\\
        \hline
        \multirow{2}{*}{$>10$ $nmol$} & $a_0$ & 1.095 & 0.029 & 37.817 & 0.000\\
        &$a_1$ & 0.063 & 0.003 & 19.201 & 0.000\\
        \hline
    \end{tabular}
    \caption{The OLS results for fitting $\lambda$ via $\exp(-\mu)$ partitioned at $10$ $nmol$. Note the final column is the $p-$value associated with each coefficient. Assuming a significance level of 0.05, then only $a_1$ for $<10$ $nmol$ is not statistically significantly. This suggests that $\lambda$ is not a function of concentration when $<10$ $nmol$. }
    \label{tab:coefs}
\end{table}

In summary, the type of transport for the series can be categorized based on the OLS regression coefficients of Equation~\ref{eq:lambda_expmu}. 
More specifically, since Knudsen diffusion coefficient is not a function of concentration, molecular diffusion occurs only when $\lambda$ is a function of concentration, and a hybrid regime is obtained when both regression coefficients are statistically significant.
Table~\ref{tab:fingerprint} summarizes the fingerprints associated with a given transport based on OLS. 
Validation of the OLS model can be used through a variety of statistical measures such as the Coefficient of Determination $(R^2)$ or $p$-value associated with the regression coefficients.

\begin{table}[]
    \centering
    \begin{tabular}{p{0.33\linewidth} p{0.33\linewidth} p{0.33\linewidth}}
         Transport & OLS Formula & Indicator  \\
         \hline
         Knudsen Diffusion &  $\boldsymbol{\lambda} = a_0$  & only the intercept is significant \\
         \hline
         Non-Knudsen transport&  $\boldsymbol{\lambda} = a_0 + a_1 \exp\left(-\boldsymbol{\mu}\right)$ & both the intercept and coefficient of $\exp(-\mu)$ are statistically significant \\
         \hline
    \end{tabular}
    \caption{Fingerprints for each transport type determined via Ordinary Least Squares where $a$ indicates a regression coefficient. Based on a series of obtained estimates of each parameter, the results of the least squares fit allows for determining if the rate parameter of the transport is affected by the concentration.}
    \label{tab:fingerprint}
\end{table}

% scaled exp(-mu) = 17.011 + 5.583 with a R2 of 0.99
% scaled eta = 17.011 + 5.576 with a R2 of 0.98

\section{Conclusion}

Four main concepts were presented: the generalization of the standard diffusion curve to account for non-Knudsen diffusion; 
proposed probability distribution theory to quantitatively measure the effects of the rate of transport coefficient and concentration without any assumptions of the reactor configuration/boundary conditions;
applied the methodology to experimental data using argon for qualitatively discriminating Knudsen and non-Knudsen diffusion; 
and interpretation of the obtained parameters was performed with respect to transport.
Constancy of the rate parameter $\lambda$ indicates the experiemnt is within the Knudsen regime. 
Conversely, change of $\lambda$ with respect to concentration is indicative of the non-Knudsen regime. 
Qualitatively, the $\exp(-\mu)$ characteristic can be considered the average concentration of the injected gas quantity and does not depend on the transport regime. 
Furthermore, when in the non-Knudsen regime the rate transport coefficient was found to be proportional to the average concentration signifying the same dependence on injection size.
In the future, additional work will be performed to include data sets of varying pulse intensity were reaction is also present (rather than pure transport).

The ability to account for non-Knudsen diffusion within the TAP response allows for experiments to be designed to include gas/gas interactions. 
This is helpful in discriminating gas phase chemistry from surface chemistry for certain types of reactions, e.g., oxidative coupling of methane.
It was also shown that the GDC was able to properly estimate the corresponding distribution coefficients in the presence of a variety of instrument drift and noise. 
As such, the experimental analysis is more consistent and is completely data-driven which helps alleviate the steep learning curve of TAP data analysis. 
Future work includes extension of the GDC to include extentions for revealing transport-reaction systems, calculations of the intrinsic kinetic rate coefficient, separation of further mixtures of GDC within pump-probe data, and quantitatively measuring complex interactions between gas species within an experiment.

\section*{Acknowledgments}
This work was supported by the U.S. Department of Energy (USDOE), Office of Energy Efficiency and Renewable Energy (EERE), Industrial Efficiency and Decarbonization Office (IEDO) Next Generation R$\&$D Project DE-FOA-0002252-1175 under contract no. DE-AC07-05ID14517.

\printbibliography

\end{document}